\pdfoutput=1
\documentclass[11pt, conference]{IEEEtran}
\usepackage{graphicx}
\usepackage{cite}
\usepackage{amsmath}
\usepackage{epstopdf}
\usepackage{graphics}
\usepackage{algorithm}
\usepackage{algpseudocode}
\usepackage{listings}
\usepackage{algorithmicx}
\usepackage{multirow}
\usepackage{multicol}
\usepackage[hyphens]{url}
\usepackage[flushleft]{threeparttable}
\usepackage{array}
\usepackage{mathtools}
\usepackage{mathtools}

\DeclarePairedDelimiter\floor{\lfloor}{\rfloor}
\usepackage{pgfplots}
\pgfplotsset{width=10cm,compat=1.9}
\usepackage{filecontents}
\usepackage{amsmath} 
\usepackage{hyperref}
\usepackage{pgfplots}
\usepgfplotslibrary{fillbetween} 
\usepackage{filecontents}
\newcolumntype{P}[1]{>{\centering\arraybackslash}p{#1}}

\IEEEoverridecommandlockouts
\begin{document}

\title{Formal Methods for an Iterated Volunteer's Dilemma}

\author{Jacob Dineen, \quad A S M Ahsan-Ul Haque, \quad Matthew Bielskas\\
Department of Computer Science, University of Virginia, Charlottesville, VA 22904\\
{\tt\small [jd5ed, ah3wj, mb6xn]@virginia.edu}
}

\maketitle
\begin{abstract}
Game theory provides a framework for studying communication dynamics and emergent phenomena arising from rational agent interactions \cite{phd10-Umm}. We present a model framework for the Volunteer's Dilemma with four key contributions: (1) formulating it as a stochastic concurrent nn
n-player game, (2) developing properties to verify model correctness and reachability, (3) constructing strategy synthesis graphs to identify optimal game trajectories, and (4) analyzing parameter correlations with expected local and global rewards over finite time horizons.
\end{abstract}

\section{Introduction}
We express the Volunteer's Dilemma as a stochastic game using the PRISM Model Checker, enabling systematic parameter tuning to understand game dynamics. This approach allows us to examine how parameter changes affect expected player rewards and to derive probabilistic graphs representing optimal or near-optimal strategies.

Previous work \cite{pbg} formulated the Public Good Game as a concurrent stochastic game, evaluating optimal strategies for fixed parameters including game length and resource distribution scaling factors. Our model similarly employs a finite state representation where agents choose discrete resource portions from their initial allocation. However, it differs fundamentally as the Volunteer's Dilemma represents a collective good game, whereas the Public Good Game focuses on localized reward maximization without explicit competition or zero-sum dynamics.

To our knowledge, this is the first application of PRISM to study the Volunteer's Dilemma as an iterated game—where the environment undergoes soft resets between rounds. Our objectives are threefold: first, verify model correctness by ensuring win conditions remain achievable; second, analyze parameter correlations with expected rewards through systematic tuning; and third, examine how game iterations manifest in synthesized strategy graphs. This analysis aims to reveal subtle aspects of Volunteer's Dilemma dynamics and generate new research questions.

\section{Background}
\subsection{Game Theoretic Scenarios}
One-shot games such as the Prisoner's Dilemma are typically modeled using payoff matrices, where players choose strategies and act concurrently and independently. In contrast, extensive form games incorporate sequential mechanisms where players act after observing their predecessors' strategies and state transitions. Iterated (or repeated) games, a subset of extensive form games, examine behavior over extended or infinite time horizons. These approaches have yielded insights into behavioral economics and rational choice theory across multiple disciplines.
Stochastic games, which incorporate probabilistic dynamics, arguably best reflect real-world systems. Typically modeled in extensive form, these games reveal complex long-run behaviors and have been applied to diverse domains including social welfare and public goods provision, robot coordination, and investment and auction scenarios \cite{hauser2019social, kwiatkowska2019equilibria,santos2019equilibria, biro2013analysis}.
\subsection{The Volunteer's Dilemma}
The Volunteer's Dilemma is a concurrent multi-agent game where each agent faces two choices:
\begin{enumerate}
\item \textbf{Cooperate}: Incur a small personal cost to produce a public benefit
\item \textbf{Defect}: Free-ride, hoping others will cooperate
\end{enumerate}
Agents decide independently and simultaneously. While the incentive to free-ride exceeds the incentive to volunteer, universal defection results in collective loss. Conversely, cooperation by at least one agent benefits all participants.
\begin{table}[H]
\centering
\caption{Payoff matrix}
\renewcommand{\arraystretch}{1.35}
\begin{tabular}{|l|l|l|}
\hline

& at least one other cooperates & all others defects \\ \hline
cooperate & 0 & 0 \\ \hline
defect & 1 & -10  \\ \hline
\end{tabular}
\end{table}

Here, agents prefer defection (payoff of 1) to cooperation (payoff of 0), yet universal defection yields the worst outcome (payoff of -10 for all).
The Volunteer's Dilemma manifests in numerous real-world contexts. In meerkat colonies, sentries who watch for predators increase their own vulnerability while protecting the group. The dilemma also illuminates collective action problems in democratic systems. Consider an election where one candidate enjoys overwhelming support: individual supporters may abstain from voting, reasoning their candidate will win regardless. However, widespread adoption of this logic could lead to the candidate's defeat—a paradox of rational individual choices producing irrational collective outcomes.
\begin{figure}[h!]
\begin{center}
\includegraphics[width=0.75\columnwidth]{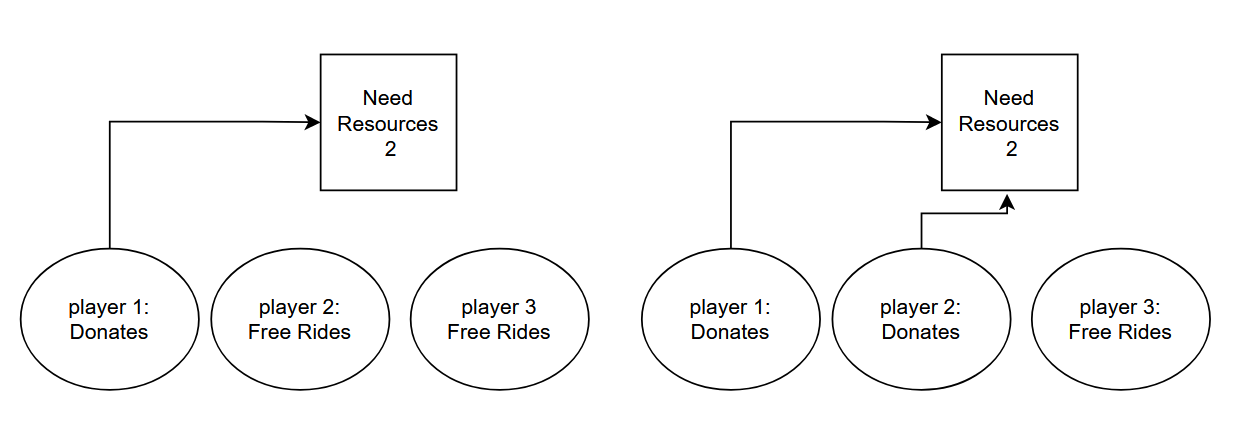}
  \caption{Volunteer's Dilemma. \textbf{Left:} We have a situation where the number of cooperating agents within the system is less than the total required resources for collective group benefit. In this case, no agent in the system benefits. \textbf{Right:} Here, the number of agents within the system who choose to cooperate is $\geq$ the total number of resources needed. All agents operating within the system, even those choosing to defect, benefit.}
  \label{fig:methods}
  \end{center}
\end{figure}

\section{Design Overview}

\textit{Concurrent stochastic multi-player games} (CSGs) extend stochastic games (SGs) from the 1950s to model group dynamics in collaborative or competitive settings where environmental states evolve through agent feedback. From any state $s \in S$, immediate rewards depend on actions taken by all agents $v \in V$. While stochastic multi-player games (SMGs) employ turn-based mechanics with individual or joint state transitions \cite{hutchison_prism-games_2013}, CSGs require simultaneous action selection. Formally, a CSG is represented as:

$$\mathrm{G}=(N, S, \vec{S}, A, \Delta, \delta, AP, L)$$

where $N$ denotes players, $S$ denotes states, $A$ denotes actions available to agent $v_i$ at time $t$, $\Delta$ is the action assignment function, $\delta$ is the probabilistic transition function, $AP$ represents atomic propositions, and $L$ is the labeling function. In CSGs, strategies resolve choice analogously to how policies resolve nondeterminism in MDPs \cite{kwiatkowska2019equilibria}.

Since environmental states and expected future payoffs depend on simultaneous agent actions, we implement our CSG using PRISM Games \cite{hutchison_prism-games_2013, KPW16, CFK+12b}, an extension of the Probabilistic Symbolic Model Checker (PRISM).

\subsection{Game Parameters}

\begin{enumerate}
    \item $k \in \{1,2,3,..., k_{max}\}$: Current episode number in the finite-length game
    \item $k_{max}$: Maximum episodes specified as environment input
    \item $V$: Agent set, where $|V| = n = 3$ (fixed for this analysis)
    \item $r_{init} = 100$: Initial resource allocation per agent per episode
    \item $c_i$: Current resources for agent $v_i$ at round $k$ (dynamically updated)
    \item $s_i \in \{0, 0.5, 1.0\} \cdot c_i$: Resources shared by agent $v_i$ at round $k$, where $s_i \leq c_i$
    \item $r_{needed}$: Fixed threshold for winning each round, where $r_{needed} < 100n$ ensures achievability with partial cooperation
\end{enumerate}

\subsection{Action Space}

To manage computational complexity, we discretize donations into three actions $A = \{a_0, a_{50}, a_{100}\}$:

\begin{table}[H]
\centering
\caption{Volunteer's Dilemma Action Space}
\resizebox{\columnwidth}{!}{%
\begin{tabular}{|p{1.5cm}|p{2.5cm}|p{7cm}|}
\hline

        variable & action  & definition \\ \hline
        
        a0 & Free Ride &  A player here chooses to contribute nothing to the pot of $r_{needed}$. This player is known in literature as a free-rider. They are hopeful that total group contribution still results in immediate payoff without sacrificing any of their resource allocation.\\ \hline
        
        a50 & Partial Contribution &  A player taking this action will contribute $\floor{(0.5 * c_i)}$ resources. \\ \hline

        a100 & Total Contribution & This action entails contribution in totality. All available resources will be pushed toward $r_{needed}$. An agent taking this action could be seen as altruistic, as they may perceive the good of the many to outweigh the good of themselves. \\ \hline
\end{tabular}
}
\end{table}

\subsection{Reward Structure}

At round $k$, agents starting in state $s_0$ concurrently select actions. Victory requires total contributions $\sum_{i=1}^{n} s_i \geq r_{needed}$ (Figure \ref{fig:reward}). Let $S^k = \sum_{i=1}^{n} s^{k}_{i'}$ denote total contributions at round $k$.

\begin{figure}[ht!]
\centering
\begin{tikzpicture}
\begin{axis}[
    title={Reward Distribution ($10^2$ resources)},
    axis lines = left,
    width=7cm,height=5cm,
    xlabel = {Over-donation: $\sum_{i=1}^{n} s_i - r_{needed}$ ($10^2$)},
    ylabel = {$R_{i}^{k}$ ($10^2$)},
]
\addplot table [x=r_donated, y=global reward, col sep=comma] {data.csv};
\addlegendentry{$R_{i}^{k}=\sum_{i=1}^{n}r^{k}_i$}
\addplot +[mark=none, dashed] coordinates {(0, 0) (0, 4)};
\addlegendentry{Optimal Strategy}
\end{axis}
\end{tikzpicture}
\caption{Reward distribution for $r_{needed} = 200$, $n = |V| = 3$, $f = 2$. The plot illustrates three distinct regions: (1) when $\sum_{i=1}^{n} s_{i'} < r_{needed}$, no rewards are distributed; (2) when $\sum_{i=1}^{n} s_{i'} = r_{needed}$, optimal joint strategy yields maximum reward with minimal resource expenditure; (3) when $\sum_{i=1}^{n} s_{i'} > r_{needed}$, rewards decay linearly according to Equation \ref{eq:1}, penalizing over-donation. Current resources at round $k$ follow the update function in Equation \ref{eq:3}.}
\label{fig:reward}
\end{figure}

The immediate reward for agent $i$ is:

\begin{equation} \label{eq:1}
r^{k}_i = \begin{cases}
0 & S^k < r_{needed} \\[3pt]
\frac{r_{needed} \cdot f}{|V|} & S^k = r_{needed} \\[3pt]
\frac{r_{needed} \cdot f - 0.014(S^k - r_{needed})}{|V|} & S^k > r_{needed}
\end{cases}
\end{equation}

\begin{equation} \label{eq:2}
R_{i}^{k} = \sum_{j=1}^{n} r^{k}_j
\end{equation}

\begin{equation} \label{eq:3}
c^{k+1}_{i} \leftarrow \min\left(r_{max}, \left\lfloor c^{k}_{i} - s^{k}_{i'} + \frac{R_{i}^{k}}{|V|} \right\rfloor\right)
\end{equation}

Here, $c_i - s_{i'}$ represents donation cost (resources retained after donation), $f$ scales rewards to prevent donor penalties relative to free-riders, and $s_{i'}$ denotes the post-transition state. Resources gained at timestep $k$ aggregate into $c^{k+1}$, enabling donation in subsequent rounds.

\begin{table}[ht]
    \centering
    \begin{tabular}{|c|c|l|l|l|l|l|l|}
    \hline$v_{i}$ & $k$ & $r_{i}^{i n i t}$ & $r_{n e e d e d}$ & $a_{i}$ & $c_{i}^{k}-s_{i^{\prime}}^{k}$ & $r_{i}^{k}$ & $c_{i^{\prime}}^{k+1}$ \\
    \hline 1 & 1 & 100 & 200 & 100 & 0 & 100 & 100 \\
    \hline 2 & 1 & 100 & 200 & 100 & 0 & 100 & 100 \\
    \hline 3 & 1 & 100 & 200 & 0 & 100 & 100 & 200 \\
    \hline
    \end{tabular}
    
        \label{tab:notoverdonation}
    \caption{No Over Donation + WIN}

\end{table}

\begin{table}[ht]
    \centering
    \begin{tabular}{|l|l|l|l|l|l|l|l|}
    \hline$v_{i}$ & $k$ & $r_{i}^{i n i t}$ & $r_{n e e d e d}$ & $a_{i}$ & $c_{i}^{k}-s_{i^{\prime}}^{k}$ & $r_{i}^{k}$ & $c_{i^{\prime}}^{k+1}$ \\
    \hline 1 & $\mathrm{k}$ & 0 & 200 & 0 & 0 & 57 & 57 \\
    \hline 2 & $\mathrm{k}$ & 500 & 200 & 250 & 250 & 57 & 307 \\
    \hline 3 & $\mathrm{k}$ & 200 & 200 & 100 & 100 & 57 & 157 \\
    \hline
    \end{tabular}
    \caption{Over-donation + WIN (Decayed Reward)}
    \label{tab:overdonation}
\end{table}

{Mock Gameplay:  Consider a simple model with three players. Table~\ref{tab:notoverdonation} shows the initial run through the CSG. The players transition through the system perfectly and gain max possible global and local rewards. The total resources after a WIN condition are perfectly met are greater than when the round started. In Table~\ref{tab:overdonation} at the $kth$ round, players have gained resources beyond their initial allocation. A player can donate ($p_{a_{i}}$ * $c_i$) resources at this step. Players 2 and 3 donate half their resources. They have over-donated, and the reward passed back is less than optimal.}

Following psychological studies on altruism \cite{altruism}, we penalize over-donation to account for ulterior motivations. When $S^k > r_{needed}$, rewards decay linearly according to Equation \ref{eq:1}, as illustrated in Figure \ref{fig:reward}.

Unlike traditional static games with complete environmental resets between rounds, our model allows resource accumulation: rewards from round $k$ become available capital at round $k+1$. This transforms the binary volunteer decision into a continuous donation choice, enabling analysis of long-run behavioral dynamics through strategy synthesis (Figure \ref{fig:methods}).

\section{Experiments and Results}

For static games, state space size follows $|S| = n^{|A|}$. However, our dynamic winning conditions depend on current game state, expanding possible joint policies leading to WIN/SAT conditions as resources accumulate through reward feedback. Consequently, state space grows exponentially over time. In the first round with $r_{init} = 1$, only $\binom{n}{r_{needed}}$ transitions yield perfect WIN conditions without over-donation decay. As $c_i \rightarrow r_{max}$, $|S|$ increases according to $|S| = 1.6978 e^{3.0479k}$ for parameters $\{k_{max} = 4, r_{init} = 100, r_{needed} = 200, n = 3\}$. We constrain $k_{max} \leq 4$ since extending to 5 and 6 rounds yields approximately 7 million and 148 million states, respectively.

\subsection{Model Correctness}

We verify model correctness using fixed parameters: $|V| = 3$ agents, $r_{init} = 100$ initial resources, $r_{max} = 1000$ local resource threshold, and $k_{max} = 4$ maximum rounds. Properties are formulated using rPATL, which combines PCTL and ATL \cite{rpatl}.

To ensure proper reward accrual, we verify that agents can eventually achieve $c_i \geq r_{init}$ with nonzero probability after $k$ rounds, indicating successful winning conditions. The piecewise reward function ensures $c_i^k < c_i^{k+1}$ when this condition is unmet. However, if $\sum_{i=1}^{n} c_i < r_{needed}$ at any point, this property becomes unsatisfiable. 

While PRISM Games lacks CTL operator support, preventing direct verification of $E[F \text{ good}]$ where $\text{good} = \sum_{i=1}^{n} c_i > r_{needed}$, we circumvent this by requiring $2 \cdot r_{needed}$ resources—achievable only after the first round:

\begin{equation}\label{eq:4}
\begin{aligned}
\text{good} &= \sum_{i=1}^{n} c_{i} > 2 \cdot r_{\text{needed}} \\
&<<p_1, p_2, p_3>> P_{\geq 1.0}\left[F_{\leq k_{\max}+1} \text{"good"}\right]
\end{aligned}
\end{equation}

In rPATL, the $<<C>>$ operator denotes player coalitions \cite{rpatl}. For our cooperative game, all players form a single coalition maximizing expected reward. Property \ref{eq:4} asserts existence of a joint strategy achieving "good" within $k_{max}$ steps with probability 1.0. This evaluates FALSE in round one but TRUE for subsequent rounds through $k_{max} = 4$, confirming model viability. Table III details probabilistic reachability through PRISM Games GUI. As gameplay progresses with round-wise satisfaction, satisfiable states increase due to resource accumulation enabling more donation combinations yielding rewards.

\begin{table} \label{table:3}
    \centering
        \caption{VGD Probabilistic Reachability Analysis}

    \begin{tabular}{|l|l|l|l|l|l|}
    \hline
        Round & States & Y & N  & M & Y / (Y + N) \\ \hline
        1 & 2 (1 init) & 0 & 2 & 0 & 0\% \\ \hline
        2 & 55 (1 init) & 6 & 48 & 1 & 11.1\% \\ \hline
        3 & 1162 (1 init) & 141 & 1009 & 12 & 12.3\% \\ \hline
        4 & 27065 (1 init) & 2724 & 8766 & 85 & 23.7\%\\ \hline
    \end{tabular}
\end{table}

\subsection{Property Verification}

Following model validation, we construct properties for CSG reachability analysis. Table III's probabilistic reachability data (Yes/No/Maybe) appears in PRISM logs rather than direct property verification. For probability-based properties, Boolean results indicate whether at least one model state satisfies the property (corresponding to at least one "Yes"). For optimization properties, results return extremal values while (Y,N,M) data becomes irrelevant. 

Let $R_i^{total} = \sum_{k=1}^{k_{max}} r_i^k$ denote cumulative reward for player $i$. We present several property templates:

\begin{equation}
<<p_1,p_2,p_3>>R\{R_1^{total}\}\max=?[F \ k=k_{\max}+1]
\end{equation}

This returns Player 1's maximum cumulative reward after $k_{max}$ rounds.

\begin{equation}
\begin{aligned}
<<p_1:p_2,p_3>>\max=? &\left(R_1^{total}[F \ k=k_{\max}+1]\right. \\
&\left.+ \sum_{j=2,3} R_j^{total}[F \ k=k_{\max}+1]\right)
\end{aligned}
\end{equation}

Here Player 1 opposes Players 2 and 3, forming two coalitions. The returned value maximizes when coalitions independently optimize rewards, where $\sum_{j=2,3} R_j^{total}$ represents the combined reward for the opposing coalition.

\begin{equation}
<<p_1,p_2,p_3>>P^{\geq 1}[F \ \sum_{i=1}^{n} c_i < 200]
\end{equation}

This probability-based property returns 1 if a state where total resources fall below 200 remains reachable. PRISM logs reveal the fraction of states satisfying this inequality.

\begin{equation}
<<p_1,p_2,p_3>>P_{\max}=?[F_{\leq k_{\max}+1} \ c_1<c_2]
\end{equation}

This returns maximum probability of Player 2 surpassing Player 1's resources after $k_{max}$ rounds. Expected value is 1 since our CSG imposes no inter-player resource constraints, with minimum probability zero.

\begin{figure*}[ht!]
\begin{center}
\includegraphics[width=1.75\columnwidth]{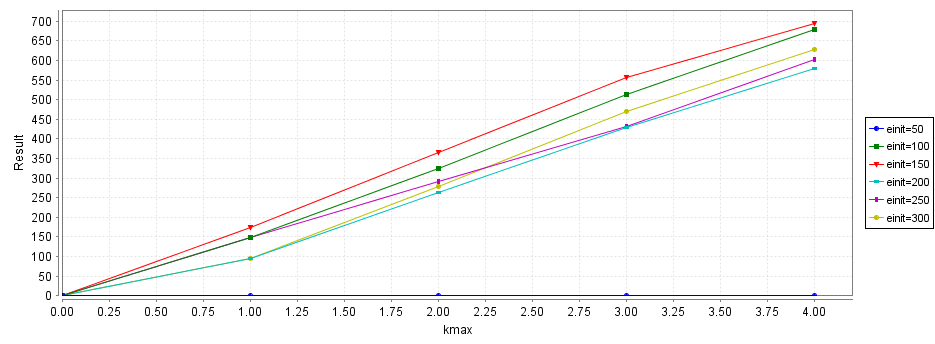}
  \caption{An iterated run through the system with variable initial resources. The y-axis represents the total, aggregate group reward through time k. The different lines represent varying initial state conditions.  }
  \label{fig:rewardmax}
  \end{center}
\end{figure*}

\subsection{Reward Maximization}

We analyze global reward maximization using the property:
\begin{equation}
<<p_1,p_2,p_3>>R\left\{\sum_{i=1}^{n} R_i^{total}\right\}\max=?[F \ k=k_{max}+1]
\end{equation}

where $\sum_{i=1}^{n} R_i^{total}$ represents total system rewards accumulated through round $k$. Figure \ref{fig:rewardmax} reveals a paradoxical finding: lower initial resource allocations yield higher maximum rewards at round 4's conclusion. Systems with $r_{init} < 200$ exhibit steeper reward growth rates and reduced stabilization across rounds. Since resource updates account for expenditures, this suggests free-riding becomes more prevalent under resource scarcity.

\begin{figure}[ht!]
\begin{center}
\includegraphics[width=1.0\columnwidth]{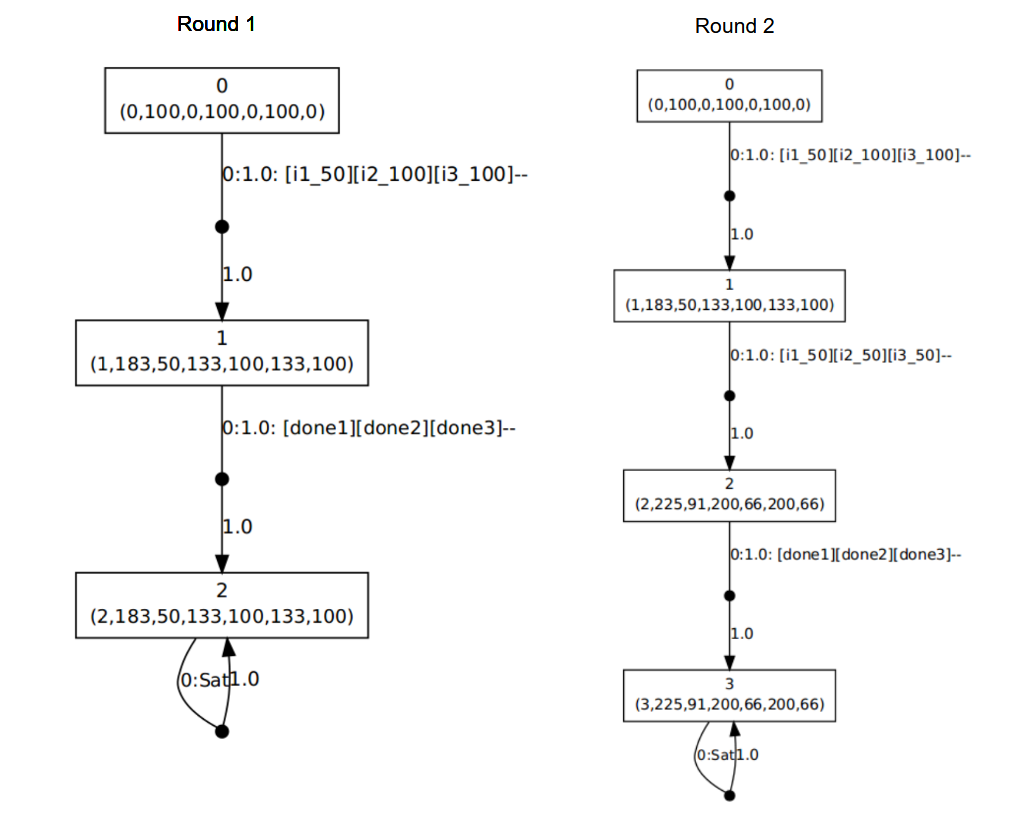}
  \caption{Strategy Graphs can be used to find an optimal controller given a property. Here, we consider $<<p1:p2,p3>>R{``r1"}max=? [ F k=kmax+1 ]$ under the specified parameter set noted above. The graphs can be read via [k, c1, s1, c2, s2, c3, s3], where branching is determined by the actions taken concurrently by all agents in the system. Some interesting patterns emerge when looking at global reward maximization against optimal strategies. On the left, results are shown for a single round. From the init state of the game, the optimal strategy is for two players to donate in totality, and for one player to partially donate. On the right, we extend this to round 2. Here, global reward maximization is achieved as a result of full participation via partial contribution. In both cases, no agent within the system freerides.}
  \label{fig:strat}
  \end{center}
\end{figure}

Optimal strategies remain unattainable round-by-round, as aggregate rewards consistently fall below the theoretical ceiling of $300n$. Figure \ref{fig:strat} illustrates this through non-intuitive strategy optimality across two gameplay rounds with $r_{needed} = 200$. We hypothesize that group optimality requires universal agent contribution.

\section{Limitations and Future Work}

\subsection{Technical Limitations}

\textbf{Reward Properties:} While PRISM supports formulating properties with extremal values over linear reward combinations, it lacks support for probability bounds (max/min) or inequalities ($P \geq p$) for such formulas. In our implementation, all rewards take the form $c_i - r_{init}$ where $c_i$ represents player resource variables, allowing direct substitution when necessary.

\textbf{Multi-Coalition Analysis:} PRISM's CSG support remains in beta, with the final release potentially imposing limitations to prevent computational intractability. Currently, properties cannot partition players into more than two coalitions when maximizing reward sums. While multiple players can appear in a single property, analyzing scenarios where each player forms an independent coalition remains infeasible. We address this by extracting maximum information from single and two-coalition properties.

\textbf{Strategy Graph Scalability:} Strategy graph analysis becomes computationally prohibitive as state space grows exponentially with gameplay progression. While single-round strategy synthesis for three players remains tractable (Figure \ref{fig:strat}), value iteration becomes infeasible as rounds increase due to state space explosion.

\subsection{Future Directions}

\textbf{Cyberphysical Applications:} The Volunteer's Dilemma framework extends naturally to free market and democratic systems. We envision applications in proximity-based cyberphysical systems for optimal route planning and traffic management, exemplified by platforms like Waze \cite{noerkaisar2016adoption, 458716}. These systems depend on voluntary real-time data sharing by "guinea pig" users (analogous to cooperators) while others exploit this information without contributing (defectors). 

This paradigm presents unique dynamics where users simultaneously consume and produce information. Cooperators who unknowingly enter suboptimal traffic conditions may derive intrinsic value from information sharing, while defectors maximize personal utility through information exploitation. Such frameworks offer rich opportunities for analyzing evolving geo-proximity behaviors and the emergence of cooperative equilibria in real-world systems where individual and collective interests intersect.

\section{Conclusion}

We presented a concurrent stochastic game model for analyzing optimal and sub-optimal behaviors in multi-agent systems under probabilistic dynamics. Through rPATL property verification, we validated model correctness and analyzed reward mechanisms across various parameter configurations. While our analysis primarily examined fixed parameter sets, exponential state space growth limited direct synthesis of collaborative strategies for long-run reward optimization.

Our cooperative game framework assumes aligned agent incentives, though real-world systems often exhibit non-cooperative dynamics characteristic of the free-rider problem. Future work should investigate scenarios where agents derive utility from minimizing opposing coalition rewards, particularly in democratic voting contexts. Such adversarial dynamics could be modeled through coalition partitioning in graphical dynamical systems, enabling analysis of combative multi-coalition interactions where strategic opposition, rather than cooperation, drives agent behavior. This extension would provide insights into equilibria emerging from competitive rather than collaborative incentive structures, better reflecting the complex motivations underlying real-world collective action problems.

{\small
\bibliographystyle{IEEEtran}
\bibliography{main}
}

\end{document}